\documentclass[
aps,%
12pt,%
final,%
notitlepage,%
oneside,%
onecolumn,%
nobibnotes,%
nofootinbib,
superscriptaddress,%
noshowpacs,%
centertags]%
{revtex4}

\begin{document}
%
\title{Off mass shell dual amplitude with Mandelstam analyticity}
\author{\firstname{Volodymyr} \surname{Magas}}
\email[]{Volodymyr.Magas@ific.uv.es}
\affiliation{Departament de F\'isica Te\'orica,
Universitat de Val\'encia\\
C. Dr. Moliner 50, E-46100 Burjassot (Val\'encia), Spain}
%
%
\begin{abstract}
A model for the $Q^2$-dependent  dual amplitude
with Mandelstam analyticity (DAMA) is proposed. The modified
DAMA (M-DAMA) preserves all the attractive
properties of DAMA, such as its pole structure and Regge asymptotics, and
leads to a generalized
dual amplitude $A(s,t,Q^2)$. This generalized amplitude
can be checked in the known kinematical limits, i.e. it should
reduce to the 
ordinary dual amplitude on mass shell, and to the nuclear structure function when $t=0$.
In such a way we complete a
unified "two-dimensionally dual" picture of strong interaction
\cite{JM0dama,JM1dama,JMfitnospin,JMfitspin}.
 By comparing the structure function $F_2$, resulting from
 M-DAMA, with  phenomenological parameterizations,
  we fix the $Q^2$-dependence in M-DAMA.
 In all studied regions, i.e. in the large and low $x$
limits as well as in the resonance region, the results of M-DAMA are in
qualitative agreement with the experiment.
\end{abstract}
\maketitle

\section{Introduction}
\label{int}

\begin{figure}[t!]
\setcaptionmargin{5mm}
\onelinecaptionstrue  
\includegraphics[height=5.0cm]{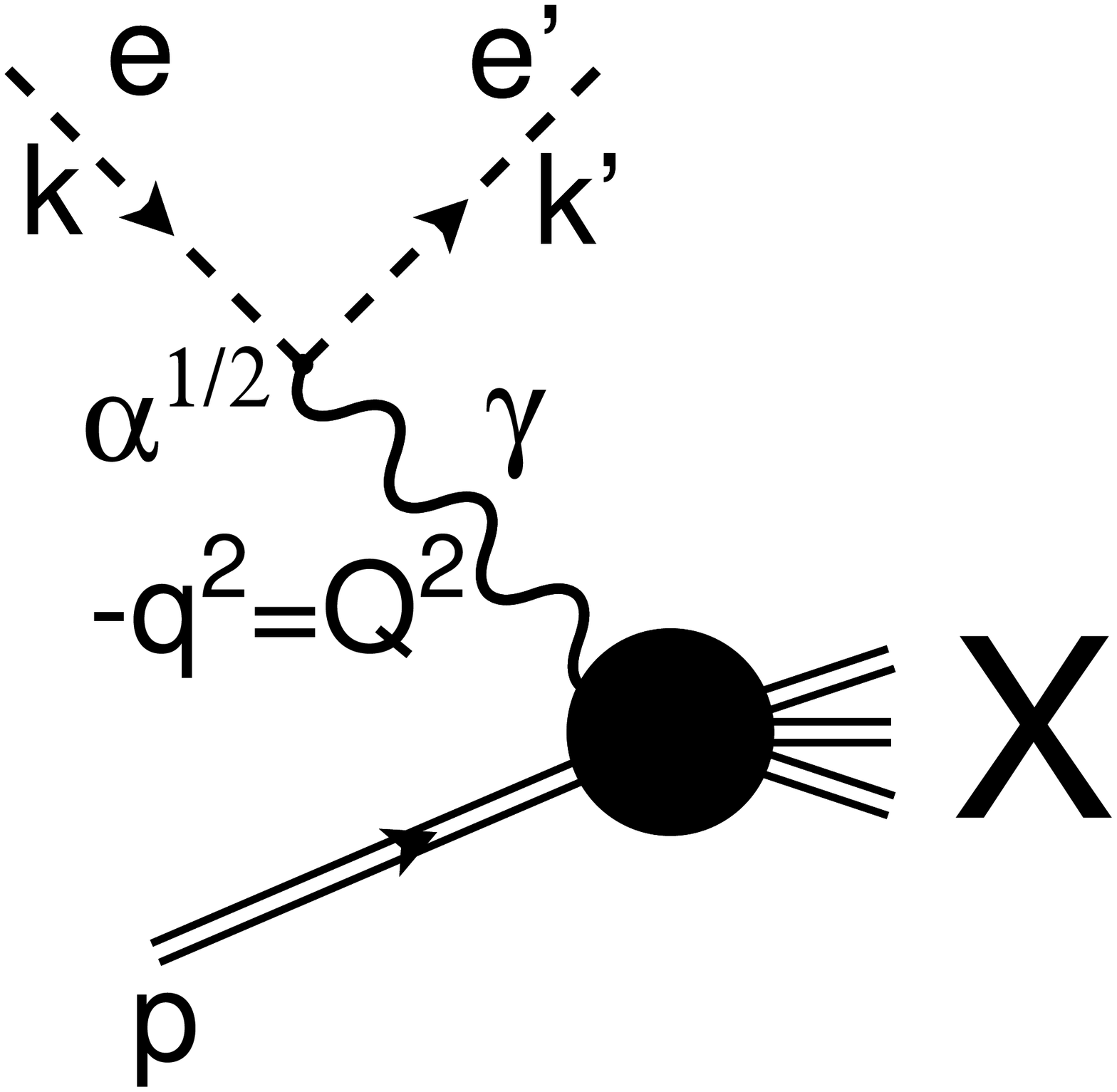}
\captionstyle{normal}
\caption{Kinematics of deep inelastic scattering.}
\label{rat}
\end{figure}

About thirty years ago Bloom and Gilman \cite{BG}
observed that the prominent resonances in  inelastic
electron-proton scattering (see Fig. \ref{rat}) do not disappear with 
increasing photon virtuality $Q^2$
with respect to the "background" but instead fall at roughly the
same rate as background. Furthermore, the smooth scaling limit
proved to be an accurate average over resonance bumps seen at
lower $Q^2$ and $s$, this is so called Bloom-Gilman or hadron-parton duality. 

For the inclusive $e^-p$ reaction we introduce virtuality $Q^2$,
$Q^2=-q^2=-(k-k')^2 \ge 0$, and Bjorken variable $x$. These variables
$x$,  $Q^2$ and
Mandelstam variable $s$ (of the $\gamma^*p$ system),  $s=(p+q)^2$, obey the
relation:
\begin{equation}
s=Q^2(1-x)/x+m^2\,,
\label{eq2}
 \end{equation}
where $m$ is the proton mass. 

Since the discovery, the hadron-parton duality was studied in a number of papers 
\cite{carlson} and the new supporting data has come from the recent experiments 
\cite{Niculescu,Osipenko}. These studies were aimed
mainly to answer the
questions: in which way a  limited number of resonances can reproduce
the smooth scaling behaviour? The main
theoretical tools in these studies were finite energy sum rules
and perturbative QCD calculations, whenever applicable. Our aim
instead is the construction of an explicit dual model combining
direct channel resonances, Regge behaviour typical for hadrons and
scaling behaviour typical for the partonic picture. Some attempts in this
direction have already been done in Refs.
\cite{JM0dama,JM1dama,JMfitnospin,JMfitspin}, which we will discuss in more
details below.

The possibility that a limited (small) number of resonances can
build up the smooth Regge behaviour was demonstrated by means of finite
energy sum rules \cite{DHS}. Later it was confused by the presence of
an infinite number of narrow resonances in the Veneziano model
\cite{Veneziano}, which made its phenomenological application
difficult, if not impossible. Similar to the case of the
resonance-reggeon duality \cite{DHS}, the hadron-parton duality was
established \cite{BG} by means of the finite energy sum rules,
but it was not realized explicitly like the Veneziano
model (or its further modifications).

First attempts to combine resonance (Regge) behaviour with
Bjorken scaling were made \cite{DG,BEG,EM} at low energies (large
$x$), with the emphasis on the right choice of the
$Q^2$-dependence, such as to satisfy the required behaviour of
form factors, vector meson dominance (the
validity (or failure) of the (generalized) vector meson dominance
is still disputable) with the requirement of
Bjorken scaling. Similar attempts in the high-energy (low $x$)
region became popular recently stimulated by the HERA data.
These are discussed in section \ref{sf}.

Recently in a series of papers
\cite{JM0dama,JM1dama,JMfitnospin,JMfitspin} authors made attempts to
build a generalized $Q^2$-dependent dual amplitude $A(s,t,Q^2)$.
This amplitude, a function of three variables, should have correct
known limits, i.e. it should reduce to the on shell hadronic scattering
amplitude on mass shell, and to the nuclear structure
function (SF) when $t=0$. In such a way we could complete a unified
"two-dimensionally dual" picture of strong interaction
\cite{JM0dama,JM1dama,JMfitnospin,JMfitspin} - see Fig.
\ref{diag}.

In Ref. \cite{JM0dama,JM1dama} the authors tried to introduce $Q^2$-dependence in
Veneziano amplitude \cite{Veneziano} or more advanced Dual Amplitude
with Mandelstam Analyticity (DAMA) \cite{DAMA}. The  $Q^2$-dependence can be
introduced either through a $Q^2$-dependent Regge trajectory \cite{JM0dama},
leading to a problem of physical interpretation of such an object, or through  the
$g$ parameter of DAMA \cite{JM0dama,JM1dama}. This last way seems to be more realistic
\cite{JM1dama}, but it is also restricted due to the DAMA model requirement
$g>1$ \cite{DAMA}. The authors \cite{JM0dama,JM1dama,JMfitnospin,JMfitspin} relate the imaginary
part of amplitude to the total cross section and then to the nucleon SF:
$F_2(x,Q^2)\sim \sigma_{tot} \sim {\cal I}m\ A(s(x,Q^2),t=0,Q^2)$, 
 which was compared to the
experimental data (we shall discuss this chain in more details in section
\ref{sf}). In this way the low $x$ behaviour of $F_2$ prescribed a
transcendental equation for $g(Q^2)$ (see \cite{JM1dama} for more details),
which
led to $g(Q^2\rightarrow\infty)\rightarrow 0$, forbidden by DAMA
definition. Therefore, such an identification of  $g(Q^2)$ is allowed only in the
limited range of $Q^2$, as it was actually stressed by the authors.

Recently this problem was also studied in the framework of the field theory.
In Ref. \cite{FT} the off shell continuation
 of the Veneziano formula was derived
 in the Moyal star formulation of Witten's string field theory.

In the papers \cite{JMfitnospin,JMfitspin} the authors went in an
opposite direction - they built a Regge-dual model with 
$Q^2$-dependent form factors, 
inspired by the pole series expansion of DAMA, which fits the SF
data in the resonance region. The hope was to reconstruct later the
$Q^2$-dependent dual amplitude, which would lead to such an expansion.
It is
important that DAMA not only allows, but rather requires nonlinear complex Regge
trajectories \cite{DAMA}. Then the trajectory with restricted real part lead to
a limited number of resonances.

A consistent treatment of the problem requires the account for the
spin dependence. It was done in \cite{JMfitspin}, and a substantial improvement
of the fit, in comparison to the earlier works \cite{JMfitnospin} ignoring the spin
dependence, was found. Nevertheless, the applicability range of the above model
\cite{JMfitspin} is limited to the resonance region, as it was actually
discussed by the authors.
For the sake of simplicity we ignore spin dependence in this paper. Our
goal is
rather to check qualitatively the proposed new way of constructing the
"two-dimensionally dual" amplitude.

\section{Modified DAMA model}
\label{smdama}

The DAMA integral is a generalization of the integral representation of the
B-function used in the Veneziano model \cite{DAMA}
\footnote{There are several integral
representations of DAMA \cite{DAMA}, here we shall use the most common one.}:
\begin{equation}
D(s,t)=\int_0^1 {dz \biggl({z \over g}
\biggr)^{-\alpha_s(s')-1} \biggl({1-z \over
g}\biggr)^{-\alpha_t(t'')-1}}\,,
\label{dama}
\end{equation}
where $a'=a(1-z)$, $a''=az$, and $g$ is a free parameter, $g>1$,
and $\alpha_s(s)$ and $\alpha_t(t)$ stand for the Regge
trajectories in the $s-$ and $t-$channels\footnote{In Ref.
\cite{DAMA} authors use the same trajectories in $s-$ and
$t-$channels. This is easy to generalize - see for example Ref.
\cite{DAMA2}. }.

In this paper we propose a modified
 definition of DAMA (M-DAMA)
with $Q^2$-dependence \cite{MDAMA}. It also can be considered as a next step in
generalization of the Veneziano model. M-DAMA preserves the
attractive features of DAMA, such as pole decompositions in $s$
and $t$, Regge asymptotics etc., yet it gains the $Q^2$ dependent form factors, 
correct $Q^2\rightarrow \infty$ limit for $t=0$ ($F_2(x,Q^2)$ at
large $x$) etc.

The proposed M-DAMA integral reads:
\begin{equation}
D(s,t,Q^2)  =  \int_0^1 dz \biggl({z \over g}
\biggr)^{-\alpha_s(s')-\beta({Q^2}'')-1}
 \biggl({1-z \over
g}\biggr)^{-\alpha_t(t'')-\beta({Q^2}')-1}\,,
\label{mdama}
\end{equation}
where $\beta(Q^2)$ is a smooth dimensionless function of $Q^2$,
which will be specified later on from studying different regimes
of the above integral.

The on mass shell limit,
$Q^2=0$, leads to the shift of the $s-$ and $t-$channel
trajectories by a constant factor $\beta(0)$ (to be determined
later), which can be simply absorbed by the trajectories and, thus, 
M-DAMA reduces to DAMA. In the general case of the virtual particle
with mass $M$ we have to replace $Q^2$ by $(Q^2+M^2)$ in the
M-DAMA integral.

Now all the machinery developed for the DAMA model (see for example \cite{DAMA})
can be applied to the above integral. Below we shall report briefly only some of
its properties, relevant for the further discussion.

\section{Singularities in M-DAMA}
\label{cases}

The dual amplitude $D(s,t,Q^2)$ is defined by the integral
(\ref{mdama}) in the domain ${\cal R}e\
(\alpha_s(s')+\beta({Q^2}''))<0$  and ${\cal R}e\
(\alpha_t(t'')+\beta({Q^2}'))<0$. For monotonically decreasing
function ${\cal R}e\ \beta({Q^2})$ (or non-monotonic function with maximum at $Q^2=0$) and for 
increasing or constant
real parts of the trajectories the first of these equations, applied
for $0\le z\le 1$, means 
\beq 
{\cal R}e\ (\alpha_s(s)+\beta(0))<0
\,. \eeq{na1} 
Similarly, the second one leads to 
\beq 
{\cal R}e\
(\alpha_t(t)+\beta(0))<0 \,. 
\eeq{na2} 
To enable us to study the
properties of M-DAMA in the domains ${\cal R}e\
(\alpha_s(s')+\beta({Q^2}''))\ge 0$ and ${\cal R}e\
(\alpha_t(t'')+\beta({Q^2}'))\ge 0$, which are of the main interest, we
have to make an analytical continuation of M-DAMA. It can be done
in the same way as for DAMA \cite{DAMA} -
basically we need to transform the integration contour in the
complex $z$ plane in such a way that $z=0$ and $z=1$ will not be
any more the end points of integration contour, instead the
contour will run around these points on an arbitrary close
distance. The important thing here is that such a procedure will
lead to an extra factor
\begin{equation}
\left\{exp[-2\pi i (\alpha_s(s')+\beta({Q^2}''))] -1 \right\}
\left\{exp[-2\pi i (\alpha_t(t'')+\beta({Q^2}'))] -1 \right\}\nn
\end{equation}
in the denominator of the M-DAMA integrand \cite{DAMA}, which generates two
moving poles $z_n$ and $z_m$ from zeros of the
denominator\footnote{Of course, the above denominator has zeros for
$n,m=-1,-2...$ also, but, as we said above, we need to make
analytical continuation only in the region where ${\cal R}e\
(\alpha_s(s')+\beta({Q^2}''))\ge 0$ and ${\cal R}e\
(\alpha_t(t'')+\beta({Q^2}'))\ge 0$. This point is not clearly
described in \cite{DAMA} - there are no poles in DAMA, for ${\cal
R}e\ \alpha(s)<0$ (or ${\cal R}e\ \alpha(t)<0$).}: \bea
\alpha_s(s(1-z_n))+\beta(Q^2z_n)=n\quad {\rm and} \nn \\
\alpha_t(tz_m)+\beta(Q^2(1-z_m))=m, \quad n,m=0,1,2...
\eea{poles}
The motion of the poles $z_n$ and
$z_m$ with $s$, $t$ and $Q^2$ depends on the particular choice of the trajectories 
and the function $\beta(Q^2)$.
The integrand (\ref{mdama}) has also two fixed branch points at $z=0$ and
$z=1$. If the trajectories $\alpha_s(s)$, $\alpha_t(t)$ or function
$\beta(Q^2)$ have thresholds and correspondingly their own branch points, then
these also generate the branch points of the M-DAMA integrand. For example
$z_s$ generated by the threshold $s_{th}$ in $\alpha_s$ trajectory will be
given by $s(1-z_s)=s_{th}$ $\Rightarrow$ $z_s=1-s_{th}/s$. Similarly
the
threshold $Q^2_{th}$ in $\beta(Q^2)$ will generate $z_Q^1=1-Q^2_{th}/Q^2$ and
$z_Q^2=Q^2_{th}/Q^2$ branch points. In this work we are not going to discuss
the threshold behaviour of M-DAMA, but we 
assume that the trajectory $\alpha_s(s)$ has a threshold and an imaginary part above it, and
correspondingly  dual amplitude $D(s,t,Q^2)$ 
also has an imaginary part above threshold.

The singularities of the dual amplitude are generated by pinches
which occur in the collisions of the above mentioned moving and fixed
singularities of the integrand.
\begin{enumerate}
\item The collision of a moving pole $z=z_n$ with the branch point $z=0$
results in a pole at $s=s_n$, where $s_n$ is defined by
\begin{equation}
\alpha_s(s_n)+\beta(0)=n\,.
\label{case1}
\end{equation}
Please, notice the presence of an extra (in comparison to 
DAMA) term $\beta(0)$. It can be considered as a shift of the
trajectory. If $\beta(0)$ is an integer number, then the
modification is trivial.
\item The collision of a moving pole $z=z_n$ with the branch point $z=1$
results in a pole at $Q^2=Q^2_n$, defined by
\begin{equation}
\alpha_s(0)+\beta(Q^2_n)=n\,.
\label{case2}
\end{equation}
In this
sense we can think about $\beta(Q^2)$ as of a kind of
trajectory, but we do not mean that it describes real physical particles. Also
we will see later that with a proper choice of $\beta(Q^2)$ we can avoid these
unphysical poles, and  $\beta(Q^2)$
required by the low $x$ behaviour of the nucleon SF is exactly of this
type.
\item Similarly, the collision of a moving pole $z=z_m$ with the
branch point $z=1$
results in a pole at $t=t_m$, defined by
\begin{equation}
\alpha_t(t_m)+\beta(0)=m\,.
\label{case3}
\end{equation}
\item The collision of a moving pole $z=z_m$ with the branch point $z=0$
results in a pole at $Q^2=Q^2_m$, defined by
\begin{equation}
\alpha_t(0)+\beta(Q^2_m)=m\,.
\label{case4}
\end{equation}
Note that if $\alpha_s(0)=\alpha_t(0)$ the poles in $Q^2$ will be degenerate.
\end{enumerate}
Generally, since poles in $s$, $t$ and $Q^2$ arise when pairs of
different singularities collide, the amplitude is free of terms
like $\sim \frac{1}{(s-s_n)(t-t_m)}$ or $\sim
\frac{1}{(s-s_n)(Q^2-Q^2_m)}$, which would possess poles
simultaneously in two variables (similarly there are no terms
possessing the poles simultaneously in all three variables).
Although in some degenerate cases this could happen  - for
example, if $\beta(x)=\alpha_s(x)$ and $\alpha_t(0)=\alpha_s(0)$,
then we could have terms like $\sim
\frac{1}{(s-s_n)(Q^2-Q^2_n)^2}$, coming from equations
(\ref{case1},\ref{case2},\ref{case4}). For further discussion we
shall consider a non-degenerated case.

\section{Pole decompositions}
\label{Pd}
Let us consider the pinch resulting from the collision of a pole at $z=z_n$
with the branch point $z=0$. The point $z_n$ is a solution of the first
equation in system (\ref{poles}):
\begin{equation}
\alpha_s(s(1-z_n))+\beta(Q^2z_n)=n\quad n=0,1,2...
\label{p1}
\end{equation}
For $z_n\rightarrow 0$ it becomes
\begin{equation}
\alpha_s(s)-s\alpha_s'(s)z_n+\beta(0)+\beta'(0)Q^2z_n=n
\label{p1a}
\end{equation}
and so
\begin{equation}
z_n=\frac{n-\alpha_s(s)-\beta(0)}{\beta'(0)Q^2-s\alpha_s'(s)}\,.
\label{p2}
\end{equation}
We see that $z_n\rightarrow 0$, when $s\rightarrow s_n$ given by eq. (\ref{case1}).
The residue at the pole $z_n$ (see \cite{DAMA} for more
details) is equal to:
\bea
 2\pi i Res_{z_n}&  = &
  \frac{1}{\beta'(0)Q^2-s\alpha_s'(s)}
\biggl({z_n \over g}
\biggr)^{-n-1}
 \biggl({1-z_n \over
g}\biggr)^{-\alpha_t(tz_n)-\beta(Q^2(1-z_n))-1} \nn \\
& &  =\frac{g^{n+1}[\beta'(0)Q^2-s\alpha_s'(s)]^{n}}
{[n-\alpha_s(s)-\beta(0)]^{n+1}} \biggl({1-z_n \over
g}\biggr)^{-\alpha_t(tz_n)-\beta(Q^2(1-z_n))-1}\,. \eea{p3} It
contains a pole at $s=s_n$ of order $n+1$. By expanding the
non-pole cofactor in (\ref{p3}) we obtain:
\begin{equation}
\biggl({1-z_n \over
g}\biggr)^{-\alpha_t(tz_n)-\beta(Q^2(1-z_n))-1} =
\sum_{l=0}^{n}C_l(t,Q^2)z_n^l
+F_n(t,Q^2,z_n)\,,
\label{p4}
\end{equation}
where
\begin{equation}
C_l(t,Q^2)=\frac{1}{l!}\frac{d^l}{dz^l}\left[\biggl({1-z \over
g}\biggr)^{-\alpha_t(tz)-\beta(Q^2(1-z))-1}\right]_{z=0}\,,
\label{p5}
\end{equation}
\begin{equation}
\frac{F_n(t,Q^2,z)}{z^{n+1}}\rightarrow const, \quad z\rightarrow 0\,.
\label{p6}
\end{equation}
Finally, inserting (\ref{p4}) into (\ref{p3})
we end up with the following expression for the pole term:
\begin{equation}
D_{s_n}(s,t,Q^2)=g^{n+1}\sum_{l=0}^{n}\frac{[\beta'(0)Q^2-s\alpha_s'(s)]^{l}C_{n-l}(t,Q^2)}
{[n-\alpha_s(s)-\beta(0)]^{l+1}}\,.
\label{p7}
\end{equation}
Formula (\ref{p7}) shows that our $D(s,t,Q^2)$ does not
contain ancestors and that an $(n+1)$-fold pole emerge on the
$n$-th level. The crossing-symmetric term can be obtained in a
similar way by considering the case 3 from the list above.

The modifications with respect to DAMA are A) the shift of the
trajectory $\alpha_s(s)$ by the constant factor of $\beta(0)$ (we can easily remove this shift
including $\beta(0)$ into trajectory); B) the coefficients $C_{l}$ are now
$Q^2$-dependent and can be directly associated with the form factors.
The presence of the multipoles, eq. (\ref{p7}), does not contradict the
theoretical postulates.  On the other hand, they can be removed
without any harm to the dual model by means the so-called Van der
Corput neutralizer\footnote{In brief, the procedure  \cite{DAMA} is to multiply the
integrand of (\ref{mdama}) by a function $\phi(z)$, which has the following 
properties:
$$ \phi(0)=0,\ \ \ \phi(1)=1,\ \ \ \phi^n(1)=0,\ \ n=1,2,3,... $$
The function $ \phi(z)=1-exp\Biggl({-{z\over{1-z}}}\biggr), $ for
example, satisfies the above conditions.}.
This procedure  \cite{DAMA} seems to work
for M-DAMA equally well as for DAMA 
 and will result in a
 "Veneziano-like" pole structure:
\begin{equation}
D_{s_n}(s,t,Q^2)= g^{n+1}
{C_n(t,Q^2)\over{n-\alpha_s(s)-\beta(0)}}\,.
\label{eq23}
\end{equation}

The $Q^2$-pole terms can be obtained by considering cases 2 and 4
from section \ref{cases}, but, as we shall see later in
section \ref{q2p}, with our choice of $\beta(Q^2)$ we avoid $Q^2$ poles.

\section{Asymptotic properties of M-DAMA}
\label{asMD}
Let us now discuss the asymptotic properties of M-DAMA. For this
purpose we rewrite the M-DAMA expression (\ref{mdama}) in the
following way:
\begin{equation}
D(s,t,Q^2)=\int_0^1 dz e^{-W(z;s,t,Q^2)} \,,
\label{mdama2}
\end{equation}
where
\begin{equation}
W(z;s,t,Q^2) = \ln \biggl({z \over g}
\biggr) (\alpha_s(s')+\beta({Q^2}'')+1)
+\ln \biggl({1-z \over
g}\biggr) (\alpha_t(t'')+\beta({Q^2}')+1)\,.
\label{W}
\end{equation}
Below a simplified notation $W(z)$ will be used instead of
$W(z;s,t,Q^2)$.

The calculations in this section will be done through the saddle point method
and we will care only about the leading order term, although the method allows to
derive subleading terms to any order. If $z_0$ is the saddle point, then the
leading term is given by:
\begin{equation}
D(s,t,Q^2)=\sqrt{\frac{2\pi}{W''(z_0)}}e^{-W(z_0)}\,.
\label{saddle}
\end{equation}

Let us prove the Regge asymptotic behaviour of M-DAMA
($s\rightarrow \infty$, $t,Q^2=const$). First we consider the
behaviour of $D(s,t,Q^2)$ for $s\rightarrow -\infty$ and fixed $Q^2$
and $t$, such that ${\cal R}e\ (\alpha_t(t)+\beta(0))+1<0$. 
In this case analytical 
continuation is not needed. The first term of
the integrand (\ref{mdama}) is a decreasing function of 
$s$ for any $0\le z<1$; it
vanishes for $z=0$. The second term vanishes at the opposite end
of the integration region.
 As it
is easy to see, the integrand has a maximum somewhere in the
middle, i.e. a saddle point, which can be found from the
equation: 
\bed 
W'(z)  = \ln \biggl({z \over g}\biggr)
(-s\alpha'_s(s(1-z)) + Q^2\beta'(Q^2z)) + {1\over
z}(\alpha_s(s(1-z))+\beta(Q^2z)+1) 
\eed
\begin{equation}
  +\ln \biggl({1-z \over g}\biggr) (t\alpha'_t(tz)-  Q^2\beta'(Q^2(1-z)))
 - {1 \over 1-z}  (\alpha_t(tz)+\beta(Q^2(1-z))+1)
=0\,.
\label{esp}
\end{equation}
Since $t$ and $Q^2$ are constants, the saddle point approaches
$z=1$ as $s \rightarrow -\infty$. For large $|s|$ and near $z=1$
there are only two important terms in eq. (\ref{esp}), the rest can be
neglected: 
\bed
 -s\alpha'_s(s(1-z))\ln \biggl({z \over g}\biggr)-
 {1 \over 1-z}  (\alpha_t(tz)+\beta(Q^2(1-z))+1)=0\Rightarrow
\eed
\begin{equation}
1-z_0 = \frac{a}{s}+O\left({1\over s^2}\right)\,,
\label{sp}
\end{equation}
where 
\beq
a=\frac{\alpha_t(t)+\beta(0)+1}{\alpha'_s(0)\ln g}\,.
\eeq{a}
Since, we are interested now only in the leading term, we can neglect
all the corrections and write: 
\bea W''(z_0)& \approx &
-s^2\alpha''_s(0)\ln g- \left(\frac{s\alpha'_s(0)\ln
g}{\alpha_t(t)+\beta(0)+1}\right)^2(\alpha_t(t)+\beta(0)+1) = \nn
\\ & & =  s^2 \left(-\alpha''_s(0)\ln g -\frac{\alpha'_s(0) \ln
g}{a}\right)\,. 
\eea{W20}
And finally, 
 \beq
D|_{s\rightarrow -\infty}  \approx    
 -s^{\alpha_t(t)+\beta(0)}
g^{\alpha_t(t)+\alpha_s(a)+\beta(Q^2)+\beta(0)+2}
 a^{-\alpha_t(t)-\beta(0)-1} 
 \sqrt{\frac{2\pi}{-\alpha''_s(0)\ln g -\frac{\alpha'_s(0) \ln
g}{a}}}  \,. 
\eea{DW} 
Thus,
\begin{equation}
D(s,t,Q^2)\sim
s^{\alpha_t(t)+\beta(0)}g^{\beta(Q^2)} \,, \quad s\rightarrow -\infty\,.
\label{W0}
\end{equation}

Now, what happens if we enter into the physical region of the
$s$-channel? In this case we have to use the analytical
continuation of M-DAMA. Using exactly the same method as in
\cite{DAMA} it is possible to show that if the trajectory
satisfies some restriction on its increase, then the Regge
asymptotic behaviour (\ref{W0}) holds for $s\rightarrow \infty$.
Of course, $D(s,t,Q^2)$ becomes a complex function, due to complex 
trajectory $\alpha_s(s)$,
 and eq. (\ref{W0})
gives the asymptotics for both real and imaginary parts. 

Thus, in the Regge limit M-DAMA has the same asymptotic behaviour as
DAMA (except for the shift $\beta(0)$). It is more interesting to
study the new regime, which does not exist in DAMA - the limit
$Q^2\rightarrow \infty$, with constant $s$, $t$. We assume that
 $\beta(Q^2)\rightarrow - \infty$ for
$Q^2\rightarrow \infty$. From eq. (\ref{esp}) we can easily find
that in this limit $z_0=1/2$. Then,
\bea
 W''(z_0)&=& 2Q^4
\beta''(Q^2/2)+8(Q^2 \beta'(Q^2/2)-\beta(Q^2/2))\nn \\ & &
+4(s\alpha'_s(s/2)-\alpha_s(s/2)-t\alpha'_t(t/2)-\alpha_t(t/2))\ \
\ \ \ \ \ \ \  \\ & & -\ln 2g
(s^2\alpha''_s(s/2)-t^2\alpha''_t(t/2))-8\, \nn
\eea{q2}
 and
\begin{equation}
D(s,t,Q^2)|_{Q^2\rightarrow \infty}
\approx (2g)^{2\beta(Q^2/2)+\alpha_s(s/2)+\alpha_t(t/2)+2}
\sqrt{\frac{2\pi}{W''(z_0)}} \,.
\label{q3}
\end{equation}
For deep inelastic scattering (DIS), as we shall see below,
if $s$ and $t$ are fixed and $Q^2\rightarrow \infty$ then 
$u=-2Q^2\rightarrow - \infty$, as it follows
from
the kinematic relation $s+t+u=2m^2-2Q^2$.
So, we need also to study the $D(u,t,Q^2)$ term in this limit. 
If $|\alpha_u(-2Q^2)|$ is growing
slower than $|\beta(Q^2)|$ or terminates when $Q^2\rightarrow \infty$,
then the previous result (eq. (\ref{q3}), $s$ to be changed to $u=-2Q^2$) 
is still valid. We shall come back to these results in the next
section to check the proposed form of $\beta(Q^2)$.

\section{Nucleon structure function}
\label{sf}

\begin{figure*}[t!]
\setcaptionmargin{5mm}
\onelinecaptionstrue  
\includegraphics[width=15cm]{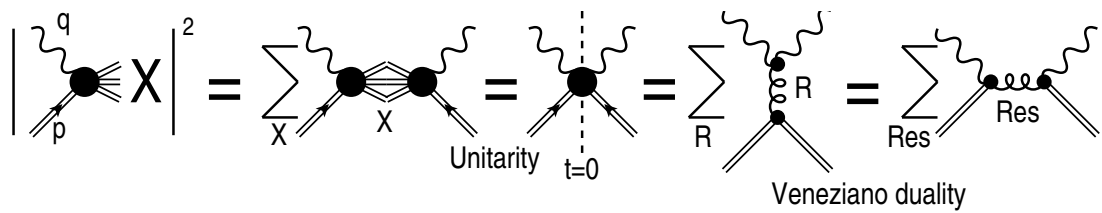}
\captionstyle{normal}
\caption{According to the Veneziano (or resonance-reggeon) duality a proper sum
of either t-channel or s-channel resonance exchanges accounts for
the whole amplitude. From \cite{JM1dama}.}
\label{d2}
\end{figure*}

The kinematics of inclusive electron-nucleon scattering,
applicable to both high energies, typical of HERA, and low
energies as at JLab, is shown in Fig. \ref{rat}.
And Fig. \ref{d2} shows how DIS is related to the forward elastic
(t=0) $\gamma^*p$ scattering, and then the latter is decomposed into a sum of the 
$s-$channel resonance exchanges. 

The total cross section is related to the SF by
\begin{equation}
F_2(x,Q^2)={Q^2(1-x)\over{4\pi \alpha (1+4m^2 x^2/{Q^2})}}
\sigma_t^{\gamma^*p}~,
\label{m23}
\end{equation}
where
$\alpha$ is the fine structure constant. In eq. (\ref{m23}) we neglected
 $R(x,Q^2)=\sigma_L(x,Q^2)/\sigma_T(x,Q^2)$, which is a reasonable
approximation.

The total cross section is related to the imaginary part of the scattering amplitude
\begin{equation}
\sigma_t^{\gamma^*p}(x,Q^2)=\frac{8\pi}{P_{CM}\sqrt{s}}\,{\cal I}m\  A(s(x,Q^2),t=0,Q^2)~.
\label{m22}
\end{equation}
where $P_{CM}$ is the center of mass momentum of the reaction,
\begin{equation}
P_{CM}=\frac{s-m^2}{2(1-x)}\sqrt{\frac{1+4m^2 x^2/{Q^2}}{s}}
\label{pcm}
\end{equation}
 for DIS.
Thus, we have 
\begin{equation}
F_2(x,Q^2)={4Q^2(1-x)^2\over{\alpha \left(s-m^2\right) (1+4m^2 x^2/{Q^2})^{3/2}}}\,{\cal I}m\  
A(s(x,Q^2),t=0,Q^2)\,.
\label{f2sigma}
\end{equation}
The minimal model for the scattering amplitude
is a sum \cite{annalen}
\begin{equation}
A(s,0,Q^2)=c(s-u)(D(s,0,Q^2)-D(u,0,Q^2)),
\label{eq34}
\end{equation}
providing the correct signature at high-energy limit, where $c$ is a
normalization coefficient ($u$ is not an independent variable, since
$s+u=2m^2-2Q^2$ or $u=-Q^2(1+x)/x+m^2$). As it was said at the beginning, we disregard
the symmetry
 properties of
the problem (spin and isospin), concentrating on its dynamics.  

In
the low $x$ limit: $x\rightarrow 0$, $t=0$, $Q^2=const$, $s=Q^2/x\rightarrow\infty$, 
$u=-s$,  we obtain, with the help of eqs. (\ref{W0},\ref{eq34}): 
\begin{equation}
{\cal I}m\  A(s,0,Q^2)|_{s\rightarrow\infty}\sim s^{\alpha_t(t)+\beta(0)+1}g^{\beta(Q^2)}\,.
\label{eq35a}
\end{equation}

Our philosophy in this section is the following: we specify a
particular choice of $\beta(Q^2)$ in the low $x$ limit
and then we use M-DAMA
integral (\ref{mdama}) to calculate the dual amplitude, and
correspondingly SF, in all kinematical domains. We
will see that the resulting SF has qualitatively
correct behaviour in all regions. Even more - our choice of
$\beta(Q^2)$ will automatically remove $Q^2$ poles.

According to the two-component duality picture \cite{FH}, both the
scattering amplitude $A$ and the structure function $F_2$ are the sums
of the diffractive and non-diffractive terms. At high energies
both terms are of the Regge type. For $\gamma^* p$ scattering only
the positive-signature exchanges are allowed. The dominant ones
are the Pomeron and  $f$ Reggeon, respectively. The relevant
scattering amplitude is as follows:
\begin{equation}
B(s,Q^2)=iR_k(Q^2)\Bigl({s\over{m^2}}\Bigr)^{\alpha_k(0)},
\label{eq4}
\end{equation}
where $\alpha_k$ and $R_k$ are Regge trajectories and
residues and $k$ stands either for the Pomeron or for the Reggeon. 
As usual, the
residue is chosen  to satisfy approximate Bjorken
scaling for the SF \cite{BGP,K}. From eqs. (\ref{f2sigma},\ref{eq4}) SF is given as:
\begin{equation}
F_2(x,Q^2)\sim Q^2
R_k(Q^2)\Bigl({s\over{m^2}}\Bigr)^{\alpha_k(0)-1}\, 
\label{eq4b}
\end{equation}
where $x=Q^2/s$ in the limit $s\rightarrow \infty$. 

It is obvious from eq. (\ref{eq4b}) that Regge asymptotics and
scaling behaviour require the residue to fall like
$\sim(Q^2)^{-\alpha_k(0)}$. Actually, it could be more involved if
we require the correct $Q^2\rightarrow 0$ limit to be respected
and the observed scaling violation (the "HERA effect") to be
included. Various models to cope with the above requirements have
been suggested \cite{R8,BGP,K}. At HERA, especially at large
$Q^2$, scaling is so badly violated that it may not be explicit
anymore.

Data show that
the Pomeron exchange leads to a rising structure function at large
$s$ (low $x$). To provide for this we have two options: either to
assume supercritical Pomeron with $\alpha_P(0)>1$ or to assume a
critical ($\alpha_P(0)=1$) dipole (or higher multipole) Pomeron
\cite{R8,wjs88,Pom1}. The latter leads to the logarithmic
behaviour
 of the SF:
\begin{equation}
F_{2,P}(x,Q^2)\sim Q^2 R_P(Q^2)\ln \Bigl({s\over{m^2}}\Bigr)\,,
\label{eq5b}
\end{equation}
which proves to be equally efficient \cite{R8,Pom1}.

Let us now come back to M-DAMA results. Using eqs.
(\ref{f2sigma},\ref{eq35a}) we obtain:
\begin{equation}
F_2\sim s^{\alpha_t(0)+\beta(0)} Q^2 g^{\beta(Q^2)}\,.
\label{n1}
\end{equation}
Choosing
\begin{equation}
\beta(0)=-1
\label{b0}
\end{equation}
we restore the asymptotics (\ref{eq4b}) and this allows us to use
trajectories in their commonly used form. It is
important to find such a $\beta(Q^2)$, which can provide for
Bjorken scaling (if one wants to take into account also the scaling
violation then the problem just gets more technical). If we choose $\beta(Q^2)$ in the form
\begin{equation}
\beta(Q^2)=d-\gamma \ln (Q^2/Q_0^2)\,,
\label{n2}
\end{equation}
with
\begin{equation}
\gamma=(\alpha_t(0)+\beta(0)+1)/\ln g=\alpha_t(0)/\ln g \,,
\label{d1}
\end{equation}
where $d$, $Q_0^2$ are some parameters, we get the exact Bjorken
scaling.

Actually, the expression (\ref{n2}) might cause problems in the
$Q^2\rightarrow 0$ limit. To avoid this, it is better to use a
modified expressions
\begin{equation}
\beta(Q^2)=\beta(0)-\gamma \ln \left(\frac{Q^2+Q_0^2}{Q_0^2}\right)=
-1-\frac{\alpha_t(0)}{\ln g} \ln \left(\frac{Q^2+Q_0^2}{Q_0^2}\right)\,.
\label{n3}
\end{equation}
This choice leads to
\begin{equation}
F_2(x,Q^2)\sim
x^{1-\alpha_t(0)}\Bigl({Q^2\over{Q^2+Q_0^2}}\Bigr)^{\alpha_t(0)}\,,
\label{eq41}
\end{equation}
where the slowly varying factor
$\Bigl({Q^2\over{Q^2+Q_0^2}}\Bigr)^{\alpha_t(0)}$ is typical for
the Bjorken scaling violation (see for example \cite{K}).

Now let us turn to the large $x$ limit. In this regime $x\rightarrow 1$,
 $s$ is
fixed, $Q^2=\frac{s-m^2}{1-x}\rightarrow \infty$ and
correspondingly $u=-2Q^2$. Using eqs.
(\ref{q3},\ref{f2sigma},\ref{eq34}) we obtain:
\begin{equation}
F_2 \sim (1-x)^2Q^4g^{2\beta(Q^2/2)}\sqrt{\frac{2\pi}{W''(z_0)}}
\left(g^{\alpha_s(s/2)}-g^{\alpha_u(-Q^2)}\right)\,.
\label{n4}
\end{equation}
For $Q^2\rightarrow \infty$ factors
 $\left(g^{\alpha_s(s/2)}-g^{\alpha_u(-Q^2)}\right)$ and 
$W''(z_0)\approx 8\gamma \ln (Q^2/Q_0^2)$ are slowly varying
functions of $Q^2$ under our assumption about $\alpha_u(-Q^2)$. 
Thus, we end up with
\begin{equation}
F_2 \sim \left(\frac{2Q_0^2}{Q^2}\right)^{2\gamma \ln 2g} \sim
(1-x)^{2\alpha_t(0)\ln 2g/\ln g}\,.
\label{n5}
\end{equation}

Let us now study $F_2$ given by M-DAMA in the resonance region.
The  existence of resonances in  SF at large
$x$ is not surprising by itself: as it follows from
(\ref{m22}) and (\ref{f2sigma}) they are the same as in $\gamma^*p$
total cross section, but in a different coordinate system.

For M-DAMA the resonances in $s$-channel are defined by the condition
(\ref{case1}). For simplicity let us assume that we performed the
Van der Corput neutralization and, thus, the pole terms appear in
the form (\ref{eq23}).
In the vicinity of the resonance $s=s_{Res}$ only the resonance
term $D_{Res}(s,0,Q^2)$ is important in the scattering amplitude and
correspondingly in the SF.

The complex pattern of the nucleon structure function in the
resonance region was developed long time ago (see, for example
\cite{Stein}). There are several dozens of resonances in the
$\gamma^* p$ system in the region above pion-nucleon threshold,
but only a few of them can be identified more or less
unambiguously for various reasons. Therefore, instead of
identifying each resonance, phenomenogists frequently considers a
few maxima (usually 3) above the elastic scattering peak,
corresponding to some ``effective'' resonance contributions. In
the Regge-dual model \cite{JMfitnospin,JMfitspin} it was shown
that for a reasonable fit it is enough to take into account three
resonance terms, corresponding to "effective"\footnote{By
"effective" trajectory the authors mean that in the fitting procedure the
parameters of these trajectories were allowed to differ from
their values at the physical trajectories. In this way the authors tried
to account for the contributions from the other resonances. The
"effective" trajectories did not move far from the physical ones,
giving thus aposteriory justification for this approach.} $\Delta$, $N$, $N^*$
trajectories with one resonance on each, plus the background. As it was
already discussed in the introduction, in the Regge-dual model the
$Q^2$-dependence was introduced "by hands". Let us now check what
we get from M-DAMA.

Using  $\beta(Q^2)$ in the form (\ref{n3}), which gives
Bjorken scaling at large $s$, we obtain from eq. (\ref{p5}):
\begin{equation}
C_1(Q^2) =
\left(\frac{gQ_0^2}{Q^2+Q_0^2}\right)^{\alpha_t(0)}
 \left[\alpha_t(0)+\ln g \frac{Q^2}{Q^2+Q_0^2} - \frac{\alpha_t(0)}{\ln g}
\ln \left(\frac{Q^2+Q_0^2}{Q_0^2}\right)\right]\,.
\label{c1}
\end{equation}
The term $\left(\frac{Q_0^2}{Q^2+Q_0^2}\right)^{\alpha_t(0)}$ gives
the typical $Q^2$-dependence for the form factor (the rest is a slowly varying function of
$Q^2$).

If we calculate higher orders of $C_n$ for subleading resonances,
we will see that the $Q^2$-dependence is still defined by the same
factor $\left(\frac{Q_0^2}{Q^2+Q_0^2}\right)^{\alpha_t(0)}$. Here
comes the important difference from the Regge-dual model
\cite{JMfitnospin,JMfitspin} motivated by introducing
$Q^2$-dependence through the parameter $g$. As we see from eq.
(\ref{eq23}), $g$ enters with different powers for different
resonances on one trajectory - the powers are increasing with the
step 2. Thus, if
$g\sim\left(\frac{Q_0^2}{Q^2+Q_0^2}\right)^{\dlt}$, then the
form factor for the first resonance is ($n=0$)
$\sim\left(\frac{Q_0^2}{Q^2+Q_0^2}\right)^{\dlt}$, and for the
second one ($n=2$) it is
$\sim\left(\frac{Q_0^2}{Q^2+Q_0^2}\right)^{3\dlt}$ etc. As
discussed in \cite{JMfitspin} the present accuracy of the data does
not allow to discriminate between the constant powers of form
factor
 (for example
Refs. \cite{Stein,Niculescu,Osipenko,DS}, and this work)
and increasing ones.

\section{How to avoid $Q^2$ poles?}
\label{q2p}
General study of the M-DAMA integral allows the $Q^2$ poles (see cases 2,4 in 
section \ref{cases}), which would be unphysical. The appearance and properties of 
these singularities depend on
the particular choice of the function $\beta(Q^2)$, and for 
our choice, given by eq. (\ref{n3}), the $Q^2$ poles can be avoided. 

We have chosen $\beta(Q^2)$ to be a decreasing function, then, according to
conditions (\ref{case2},\ref{case4}), there are no $Q^2$ poles in M-DAMA in
the physical domain $Q^2\ge 0$, if
\begin{equation}
{\cal R}e\ \beta(0)<-\alpha_s(0)\,, \quad {\cal R}e\
\beta(0)<-\alpha_t(0)\,.
\label{bt}
\end{equation}
We have already fixed $\beta(0)=-1$, eq. (\ref{b0}), and, thus, we
see that indeed we do not have $Q^2$ poles, except for the case of
supercritical Pomeron with the intercept $\alpha_P(0)>1$. Such a
supercritical Pomeron would generate one unphysical pole at $Q^2=Q^2_{pole}$
defined by equation
\begin{equation}
-1-{\alpha_P(0)\over \ln g}\ln
\left(\frac{Q^2+Q_0^2}{Q_0^2}\right)+\alpha_P(0)=0\quad
\Rightarrow \quad Q^2_{pole}=Q_0^2(g^{{\alpha_P(0)-1} \over
\alpha_P(0)}-1)\,. \label{Qp}
\end{equation}
Therefore we can conclude that M-DAMA does
not allow a supercritical trajectory - what is good from the theoretical
point of view, since such a trajectory violates the Froissart-Martin
limit \cite{Frois}.

As it was discussed above 
there are other phenomenological models which use dipole Pomeron with the
intercept $\alpha_P(0)=1$ and also fit the data (see for example
\cite{R8}).
This is a very interesting case - ($\alpha_t(0)=1$) - for the
proposed model. At the first glance it seems that we should 
anyway have a pole at $Q^2=0$. It should 
result from the collision of
the moving pole $z=z_0$ with the branch point $z=0$, where 
$\alpha_t(0)+\beta(Q^2(1-z_0))=0$ in our case. Then, checking the conditions
for such a collision: 
$$
\alpha_t(0)-t\,\alpha_t'(0)z_0+\beta(Q^2)-\beta'(Q^2)Q^2z_0=0\ \Rightarrow \ 
z_0=\frac{-\alpha_t(0)-\beta(Q^2)}{t\,\alpha_t'(0)-Q^2\beta'(Q^2)}\,,
$$
we see
that for $t=0$ and for $\beta(Q^2)$ given by eq. (\ref{n3}) the
collision is simply impossible, because $z_0(Q^2)$ does not tend to $0$
for $Q^2\rightarrow 0$. Thus, for the Pomeron with
$\alpha_P(0)=1$ M-DAMA does not contain any unphysical
singularity.

On the other hand, a Pomeron trajectory with $\alpha_P(0)=1$ does
not produce rising SF (\ref{eq4b}), as required by the experiment.
So, we need a harder singularity and the simplest one is a dipole
Pomeron. A dipole Pomeron produces poles of the second power: 
\beq
D_{dipole}(s,t_{m})\propto
\frac{C(s)}{(m-\alpha_P(t)+1)^2}\,,
\eeq{dp1} 
usually the
simple pole is also taken into account (we write a sum of simple
pole and dipole) - see for example ref. \cite{wjs88} and
references therein. Formally such a dipole Pomeron can be written
as $$ \frac{\partial}{\partial \alpha_P}
\frac{C(s)}{(m-\alpha_P(t)+1)}\,,$$ and generalizing
this  \beq D_{dipole}(s,t)= \frac{\partial}{\partial \alpha_P}
D(s,t)\,, \eeq{dp2} where $D(s,t)$ can be given for example by
DAMA or M-DAMA. Applying this expression to the asymptotic formula
of M-DAMA, eq. (\ref{W0}), we obtain a term $g^{\beta(Q^2)}
s^{\alpha_t(t)+\beta(0)} \ln s$, which then leads to a
logarithmically rising SF (for $\alpha_P(0)+\beta(0)=0$) - the 
one given  by eq. (\ref{eq5b}).

For $\beta(Q^2)$ in the form (\ref{n3}) M-DAMA will generate 
an infinite number of the $Q^2$ poles concentrated near the "ionization point" 
$Q^2=-Q^2_0$. Although these are in the
unphysical region of negative $Q^2$, such a feature of the model\\
 A) makes us 
think about $\beta(Q^2)$ as about a kind of trajectory, 
what is not the case, as it was stressed above,
 and\\ B)
might create a problem for a general theoretical treatment, for example for 
making analytical continuation in $Q^2$. To avoid this we can redefine 
$\beta(Q^2)$ in the nonphysical $Q^2$ region, for example in the following way:
\begin{equation}
\beta(Q^2)=\left\{ 
\begin{array}{l}
-1-\frac{\alpha_t(0)}{\ln g} \ln 
\left(\frac{Q^2+Q_0^2}{Q_0^2}\right)\,,
\quad {\rm for}\ Q^2\ge 0\,, \\
-1-\frac{\alpha_t(0)}{\ln g} \ln 
\left(\frac{Q_0^2-Q^2}{Q_0^2}\right)\,,  
\quad {\rm for}\ Q^2< 0\,.
\end{array}
\right.
\label{n3b}
\end{equation}
This function has a maximum at $Q^2=0$, $\beta(0)=-1$.
M-DAMA with $\beta(Q^2)$ given by eq. (\ref{n3b})
preserves all its good properties, discussed above, and 
does not contain any 
singularity in $Q^2$ (except for the supercritical Pomeron case, which we do not allow).

\section{Conclusions}
\label{concl}
\begin{figure}[t!]
\setcaptionmargin{5mm}
\onelinecaptionsfalse 
\includegraphics[width=15cm]{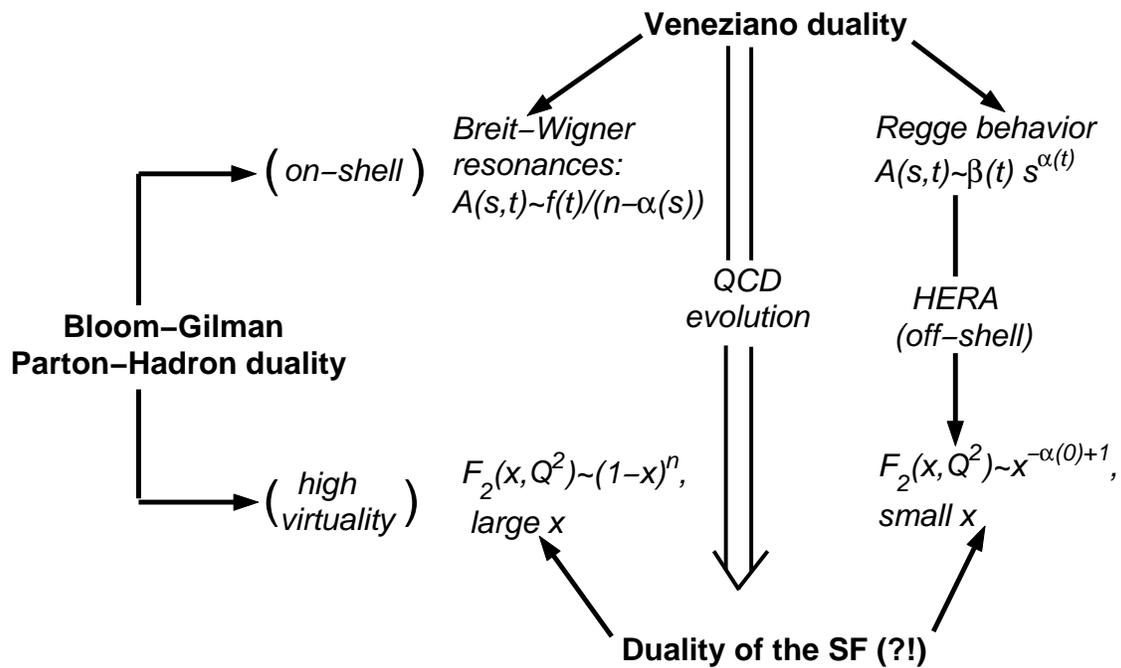}
\captionstyle{normal}
\caption{Veneziano, or resonance-reggeon
duality \cite{Veneziano} and
Bloom-Gilman, or hadron-parton duality \cite{BG}
 in strong interactions. From \cite{JM1dama}.}
\label{diag}
\end{figure}

A new model for the $Q^2$-dependent dual amplitude
with Mandelstam analyticity is proposed. The M-DAMA preserves all the attractive
properties of DAMA, such as its pole structure and Regge asymptotics, but it also
leads to generalized
dual amplitude $A(s,t,Q^2)$ and in this way realizes a
unified "two-dimensionally dual" picture of strong interaction
\cite{JM0dama,JM1dama,JMfitnospin,JMfitspin} (see Fig. \ref{diag}).
This amplitude, when $t=0$, can be related
 to the nuclear structure function. In section \ref{sf}
 we compare the SF generated by
 M-DAMA with phenomenological parameterizations,
  and in this way we fix the
 function $\beta(Q^2)$, which introduces the $Q^2$-dependence in M-DAMA, eq.
 (\ref{mdama}). The conclusion is that for both large and low $x$
limits as well as for the resonance region the results of M-DAMA are in
qualitative agreement with the experiment.

General study of the M-DAMA integral tells us about the
possibility to have poles in $Q^2$. These singularities may be
avoided with our choice of $\beta(Q^2)$, and also by putting
restriction on the physical trajectories - the use of
supercritical trajectory would lead to one $Q^2$ pole.

In the proposed formulation a $Q^2$-dependence is introduced into
DAMA through the additional function $\beta(Q^2)$. Although in the
integrand this function stands next to Regge trajectories, this, as it
was stressed already, 
 does not mean that it also corresponds to
some physical particles. There is no qualitative difference
between two ways of introducing $Q^2$-dependence into DAMA:
through the $Q^2$-dependent parameter $g$, i.e. function $g(Q^2)$
\cite{JM0dama,JM1dama} or through the function $\beta(Q^2)$. On
the other hand the second way, i.e.
 M-DAMA, is applicable for all range of $Q^2$ and it results into physically
 correct behaviour in all tested limits.

\begin{acknowledgments}
I thank L.L. Jenkovszky for fruitful and enlightening discussions.
Also, I acknowledge the support by INTAS under Grant 00-00366.
\end{acknowledgments}

\newpage
%
%
%
%
%
%
\newpage
%
\newpage
%
\newpage
%
%
%
%
%
%

\end{document}